\documentclass[fleqn,usenatbib,useAMS]{mnras}

\usepackage{graphicx}	
\usepackage{amsmath}	
\usepackage{amssymb}	
\usepackage{multicol}        
\usepackage{bm}		
\usepackage{pdflscape}	
\usepackage{adjustbox} 
\usepackage{rotating}
\usepackage{multirow}
\usepackage{array}
\usepackage[dvipsnames]{xcolor}

\newcommand{\kms}{\,km\,s$^{-1}$} 

\usepackage[T1]{fontenc}
\usepackage{ae,aecompl}

\usepackage{newtxtext,newtxmath}
\usepackage[normalem]{ulem}

\title[H$_{2}$O and CO emission from MG~J0414+0534]{Smoke on the water: CO and H$_{2}$O in a circumnuclear disc around a quasar at redshift 2.64}

\author[H.~R.~Stacey, A.~Lafontaine \& J.~P.~McKean]{
H.~R.~Stacey,$^{1,2,3}$\thanks{E-mail: stacey@mpa-garching.mpg.de}
A.~Lafontaine$^{2}$ and
J.~P.~McKean$^{1,2}$
\\
$^{1}$ASTRON, Netherlands Institute for Radio Astronomy, Oude Hoogeveensedijk 4, 7991 PD, Dwingeloo, The Netherlands\\
$^{2}$Kapteyn Astronomical Institute, University of Groningen, P.O. Box 800, 9700AV Groningen, The Netherlands\\
$^{3}$Max Planck Institute for Astrophysics, Karl-Schwarzschild Str. 1, D-85748 Garching bei M\"unchen, Germany \\
}

\date{Last updated XXX; in original form XXX}

\pubyear{2019}

\begin{document}
\label{firstpage}
\pagerange{\pageref{firstpage}--\pageref{lastpage}}
\maketitle

\begin{abstract}
We present an analysis of observations with the Atacama Large (sub-)Millimetre Array (ALMA) of the 380~GHz water emission line and CO~(11--10) emission line from MG~J0414+0534, a gravitationally lensed dusty star-forming galaxy that hosts a type~1 quasar. We also present observations at 1.6~GHz with global very long baseline interferometry (VLBI) of the radio source. We confirm the previously reported detection of the 380~GHz water line and that the flux density ratio between the two merging lensed images is reversed with respect to the radio/mm continuum. We further find tentative evidence of variability in the integrated line intensity on timescales of days. We show that the 380~GHz water line has two components of emission: a disc that is around 35~pc in diameter around the quasar, and another component of emission that is offset $\sim600$~pc perpendicular to the disc that lies close to the lensing caustic. With lens modelling of the multi-wavelength data sets, we construct a model for the quasar system consisting of a circumnuclear disc of molecular gas with a size of about 60~pc bisected by radio jets extending to a distance of about 200~pc from the radio core. Our findings suggest that observations with ALMA of high-excitation molecular lines from strongly lensed quasars could allow detailed studies of AGN accretion and feedback at the cosmic peak of black hole and galaxy growth.
\end{abstract}

\begin{keywords}
masers -- galaxies: nuclei -- galaxies: jets -- gravitational lensing: strong -- quasars: general
\end{keywords}

\section{Introduction}

Water is an abundant molecule in the Universe with a large number of observable transitions from cm to mm wavelengths, whose properties can provide valuable insights into the physical conditions of the interstellar medium (ISM) of galaxies. While H$_{2}$O emission is weak from the quiescent cold ISM, it is often associated with the dense, warm (or shock-heated) ISM in star-forming regions (e.g. \citealt{vanderWerf:2011,Jarugula:2019}) or regions in the vicinity of active galactic nuclei (AGN; e.g. \citealt{Moran:1995}). Collisional excitation or pumping from a strong radiation field can lead to population inversion, which can create the conditions needed for bright maser emission to occur \citep{Neufeld:1994}. Extra-galactic H$_{2}$O masers can be produced by a large number of individual masers in an intense starburst \citep{Cernicharo:2006,Konig:2017}. However, they are more typically associated with AGN, where synchrotron emission from the radio core can penetrate the large column densities of dust and gas in the circumnuclear region, and provide sufficient continuum luminosity to reach the critical threshold for maser emission. These are termed `megamasers', having luminosities $>10^{6}$ times higher than Galactic masers (see \citealt{Lo:2005}, for review). 

AGN megamasers have been localised with very long baseline interferometry (VLBI; 6$_{16}$--5$_{23}$ transition; rest-frequency 22.23508~GHz) at cm-wavelengths to regions that are typically $<1$~pc from the radio core. These can originate from illuminated gas near or in the accretion disc or torus, where they can be used to constrain the physical environment around the black hole (e.g. \citealt{Moran:1995,Herrnstein:1998,Klockner:2003,Kuo:2018}). Studies of AGN disc masers have enabled accretion discs to be characterised as geometrically thin discs of dense molecular gas \citep{Neufeld:1995,Herrnstein:1996,Reid:2009,Kuo:2011} with evidence of substructure \citep{Humphreys:2008}. AGN megamasers can also be produced by interactions between jet plasma and dense clouds of gas, and are therefore a useful probe of jet physics \citep{Peck:2003} and can trace jet-driven outflows of molecular gas \citep{Greenhill:2003}. Observations of megamasers at high redshift are, therefore, especially of interest as they can be used to constrain the environment around AGN at the peak of cosmic star formation and AGN activity \citep{Madau:2014}. 

Thus far, only one 22~GHz H$_{2}$O megamaser has been detected at $z\gtrsim1$, from the gravitationally lensed quasar system MG~J0414+0534 ($z=2.64$; \citealt{Impellizzeri:2008,Castangia:2011}). The next highest redshift detection is from a type 2 quasar at $z=0.66$ \citep{Barvainis:2005}, despite numerous searches of other systems at $z>1$ \citep{Wilner:1999,Ivison:2006,Wagg:2009,Bennert:2009,McKean:2011}. The megamaser emission from MG~J0414+0534 has not been localised, so the origin (whether disc, jet or starburst) has, thus far, been undetermined. However, the bright synchrotron emission from the radio source and large isotropic line luminosity of $10^{3.9}$~L$_{\odot}$ strongly suggests the megamaser emission is associated with the AGN. The blueshifted, complex velocity profile, initially reported by \cite{Impellizzeri:2008}, suggested it originates in the radio jet. However, the later tentative detection of a redshifted satellite line by \cite{Castangia:2011} may point to a nuclear disc origin.

Variability of 22~GHz H$_2$O megamasers that are associated with AGN can be caused by fluctuations in the luminosity of the seed photons, the radiative pumping source that causes population inversion \citep{Neufeld:1994,Gallimore:2001}, the movement of the masing region (such as in a disc; \citealt{Moran:1995}), or from interstellar scintillation \citep{Greenhill:1997}. These variations can occur on timescales of minutes, in the case of scintillation, to months, in the case of disc rotation. It has also been suggested that microlensing by stars in foreground lensing galaxies can result in significant (extrinsic) variability of lensed 22 GHz H$_2$O megamasers at high redshift \citep*{Garsden:2011}. The 15~month monitoring campaign of MG~J0414+0534 with the Arecibo telescope that was reported by \cite{Castangia:2011} revealed variations in the isotropic line luminosity over this timescale, consistent with an AGN origin. They did not find significant evidence of a velocity drift, indicative of rotation of any maser regions in a pc-scale accretion disc. However, high-velocity disc features found in nearby AGN, where megamasers originate from the mid-line of the disc (e.g. \citealt{Bragg:2000,Yamauchi:2005,Humphreys:2008}), fall within the inferred upper limit for any velocity drift in the 22~GHz H$_2$O megamaser emission from MG~J0414+0534.

The complex energy level structure of H$_{2}$O also results in a large number of rotational transitions emitted at far-infrared (FIR) to mm wavelengths, which can be spatially resolved with the new capabilities of sub-mm/mm interferometry. These higher frequency transitions of H$_2$O are associated with both extreme star-formation and AGN activity within FIR-luminous galaxies \citep{Omont:2011,vanderWerf:2011}, and have been used to probe the physical conditions within the ISM at high redshift \citep{Gonzalez:2014}. Given the previous detection of the 22~GHz H$_2$O megamaser, observations of MG~J0414+0534 with the Atacama Large (sub)Millimetre Array (ALMA) were recently carried out by \cite{Kuo:2019}, who made a tentative detection of the 4$_{14}$--3$_{21}$ H$_{2}$O line (rest frequency 380.197~GHz). Surprisingly, this tentative detection finds the flux ratio between the two merging lensed images (A1 and A2) to be reversed relative to the radio/mm continuum and CO~(11--10) line emission \citep{Stacey:2018b}, and counter to expectations of gravitational lensing theory \citep{Kochanek:2004}. \citeauthor{Kuo:2019} account for this with three regions of megamaser emission, where one component is coincident with the quasar and two components are separated by large distances ($\sim1.5$~kpc) from the quasar.

Here, we present a re-analysis of the ALMA observations of MG~J0414+0534 reported by \citeauthor{Kuo:2019}, and use these in combination with archival 1.6 GHz VLBI data of the radio continuum and ALMA CO~(11--10) imaging to investigate the emission properties of the 380~GHz H$_2$O line. We confirm a robust detection of the 380~GHz H$_2$O line emission and the reverse in the flux ratio of the two merging lensed images with respect to the other multi-wavelength data for this object. We localise the 380~GHz H$_2$O emission relative to the CO and radio jet components, from which we derive a model for the emission from the system. In Section~\ref{section:obs} we describe the observations and data reduction of the ALMA and global VLBI data. In Section~\ref{section:results}, we report the velocity structure and variability of the 380~GHz H$_2$O line, and describe our lens modelling analysis. Finally, in Section~\ref{section:discussion} we propose a source-plane model to describe the physical origin of the different H$_2$O components and consider the implications of our results for future searches for mm-wavelength H$_{2}$O line emission in the early Universe.

We assume the \cite{Planck:2016} model of flat $\Lambda$CDM cosmology with $H_{0}= 67.8~$km\,s$^{-1}$ Mpc$^{-1}$, $\Omega_{\rm M}=0.31$ and $\Omega_{\Lambda}=0.69$. For this cosmology, 1~arcsec corresponds to 8.2~kpc at the source redshift of 2.64.

\section{Observations and data reduction}
\label{section:obs}

In this section, we summarise the observations and data reduction of the ALMA and global VLBI data used in this analysis.

\begin{figure*}
\centering
    \includegraphics[width=\textwidth]{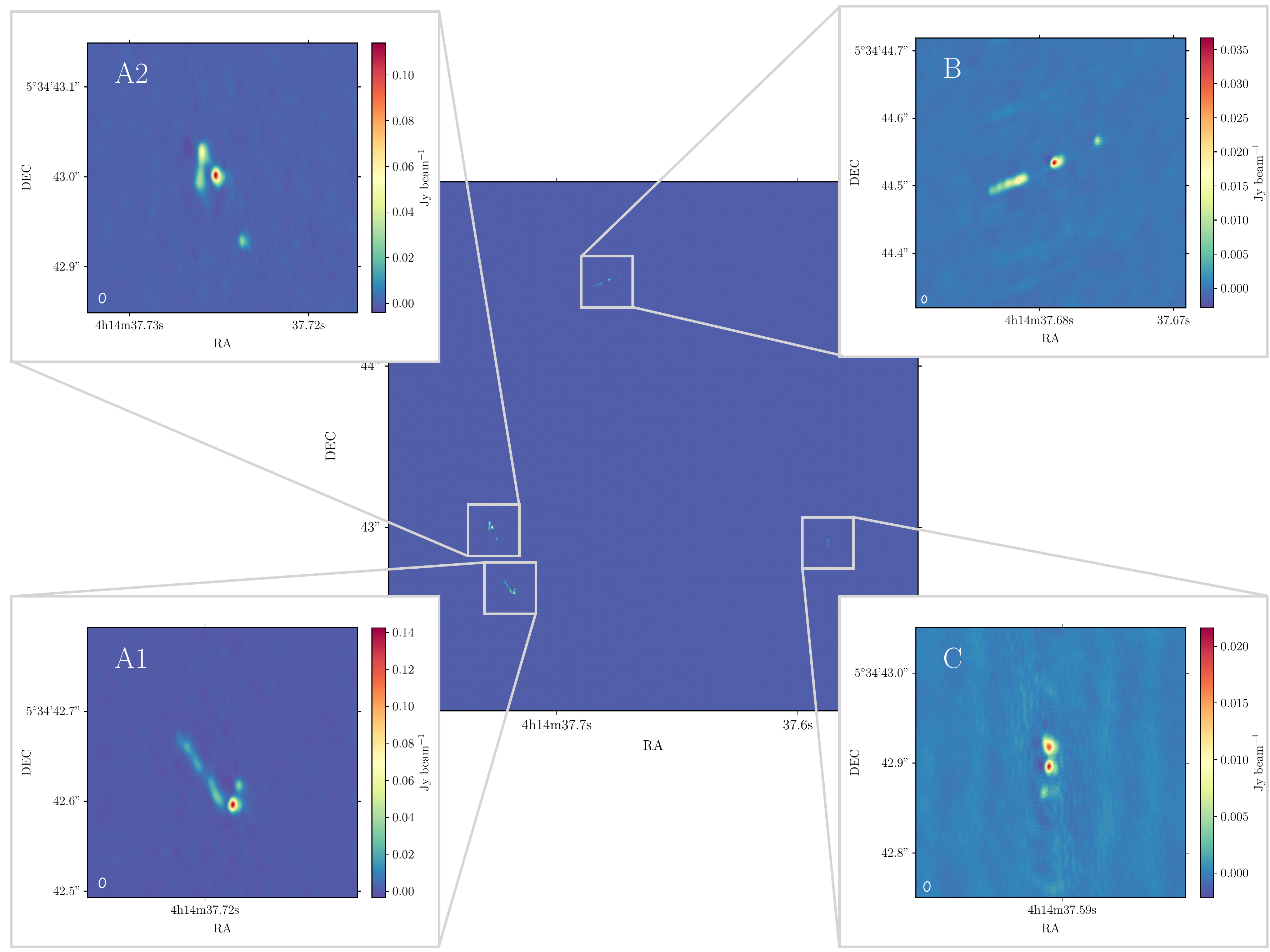}
    \caption{1.6~GHz global VLBI image of MG~J0414+0534. Inset figures show cutouts of each lensed image. The beam size is shown in the lower-left corner of the inset images and is approximately $7\times11$~mas.}
    \label{fig:0414_gvlbi}
\end{figure*}

\begin{table*}
    \caption{The radio VLBI and CO velocity components used in our lens modelling. The components are measured by fitting 2D Gaussians to the images using the task {\sc IMFIT} within CASA. The positions are defined in arcsec relative to component B-I. }
    \centering
    \adjustbox{width=\textwidth}{
    \def\arraystretch{1.3}
    \begin{tabular}{c c | c c c c | c c}
       \multicolumn{2}{c}{}  & \multicolumn{4}{c}{VLBI} & \multicolumn{2}{c}{CO} \\
         &  & I & II  & III  & IV & red & blue \\ \hline 
        \multirow{2}{*}{A1} & $\delta x$  &  $-0.589932\pm0.000005$ & $-0.60892\pm0.00004$ & $-0.58320\pm0.00002$ & $-0.63538\pm0.00010$ &  $-0.576\pm0.003$ & $-0.596\pm0.005$  \\
                         & $\delta y$ & $-1.938372\pm0.000009$ & $-1.92669\pm0.00008$ & $-1.91770\pm0.00005$ & $-1.88383\pm0.00017$ & $-1.993\pm0.003$ & $-1.863\pm0.005$ \\
        \multirow{2}{*}{A2} & $\delta x$  & $-0.723891\pm0.000006$ & $-0.73902\pm0.00001$ & $-0.69423\pm0.00003$ & $-0.74172\pm0.000016$ & $-0.739\pm0.003$ & $-0.709\pm0.005$  \\ 
                         & $\delta y$ & $-1.53326\pm0.00001$ &  $-1.50852\pm0.00004$ & $-1.60586\pm0.00005$ & $-1.53706\pm0.00008$ & $-1.520\pm0.003$ & $-1.620\pm0.005$ \\
        \multirow{2}{*}{B} &  $\delta x$ & $\equiv0.00000\pm0.00004$ & $-0.05185\pm0.00010$ & $+0.06236\pm0.00006$ & $-0.72623\pm0.00030$ & $\equiv0.000\pm0.003$ & $\equiv0.000\pm0.005$ \\ 
                         & $\delta y$ & $\equiv0.00000\pm0.00003$ & $-0.02566\pm0.00007$ & $+0.03189\pm0.00010$ & $-0.03282\pm0.00014$ & $\equiv0.000\pm0.003$ & $\equiv0.000\pm0.005$\\
        \multirow{2}{*}{C} & $\delta x$  & $+1.3549\pm0.00002$ & $+1.3549\pm0.0005$ & $+1.34980\pm0.00004$ & $+1.35486\pm0.00049$ & $+1.345\pm0.003$ & $+1.345\pm0.005$ \\  
                         & $\delta y$ & $-1.63822\pm0.00005$ & $-1.61619\pm0.00005$ & $-1.66682\pm0.00013$ & $-1.61619\pm0.00080$ & $-1.648\pm0.003$ & $-1.648\pm0.005$ \\ \hline
    \end{tabular}}
    \label{table:vlbi_comps}
\end{table*}

\subsection{1.6~GHz VLBI observations}

MG~J0414+0534 was observed with the global VLBI array on 2008 June 7, under project code GW019A (PI: Wucknitz). The observations were performed at a central frequency of 1.6~GHz in full circular polarisation (RR, LL, RL, LR). The data were correlated with a visibility integration time of 1~s, in 8 spectral windows each of 16~channels and 8~MHz bandwidth. A total of 21 antennas were used for the observation from the European VLBI Network and the Very Long Baseline Array, but only 18 antennas were found to have usable data. 

Phase referencing to nearby calibrators (J0412+0438, J0422+0219, J0409+1217) was performed at intervals of 14~min and standard fringe finder calibrators (3C286, 3C48, DA193, 0528+134) were observed at regular intervals over the 20~h observation. The data were edited, calibrated and imaged using the Astronomical Image Processing Software (AIPS). The data were manually flagged to remove radio frequency interference and off-source integrations. Observations of the fringe finders were used to remove the instrumental delays. The spectral bandpass was calibrated using observations of DA193 and 0528+134, and the relative amplitudes were set using measured system temperatures and gain curves. Observations of the phase-reference calibrator were used to determine the time-dependent phase delays and rates. Several iterations of self-calibration of the target were performed using the line-free spectral windows, down to a solution interval of 60~s, to correct for residual phase and amplitude errors. The self-calibration solutions were applied to the spectral window containing the line emission. The target was imaged and deconvolved in AIPS using a Briggs weighting of the visibilities (shown in Fig.~\ref{fig:0414_gvlbi}). The VLBI components appear structurally similar to those reported by \cite{Ros:2000} at 8.4~GHz, and previous imaging of the same data by \cite{Volino:2010}.

The task {\sc IMFIT} within the Common Astronomy Software Applications (CASA) package \citep{McMullin:2007} was used to fit 2-dimensional Gaussians to the sub-components of the four lensed images (in the image plane) to extract their positions and uncertainties (listed in Table~\ref{table:vlbi_comps}). The radio components are measured relative to the brightest component in image B, for consistency with previous studies (e.g. \citealt{Macleod:2013}). The relative image positions of the brightest radio component (sub-component I) are consistent with the relative positions of the optical emission from the quasar \citep{Falco:1997}, so this is likely emission from (or very close to) the radio core.

\subsection{ALMA observations}

The ALMA imaging of MG~J0414+0534 was taken in two distinct data sets, which we describe separately. The first, taken at 100~GHz, targeted H$_2$O, whereas the second, taken at 340~GHz, targeted high-excitation CO.

\subsubsection{100~GHz data}

MG~J0414+0534 was observed with ALMA in Band 3 under project code 2017.1.00316.S (PI: Kuo). The data were taken in four epochs on 2017 December 13, 15 (two observations) and 18 in an antenna configuration with a longest baseline of 2400~m. The data were correlated with a visibility integration time of 3~s, in four spectral windows of central frequency 92, 94, 104 and 106~GHz, respectively. The 94 and 104~GHz spectral windows covered the redshifted rest frequency of two H$_{2}$O emission lines (4$_{14}$--3$_{21}$, $\nu_{\rm rest}=380.197$~GHz, and 5$_{23}$--6$_{16}$, $\nu_{\rm rest}=336.228$~GHz). These spectral windows were each correlated with 1920 channels and 0.94~GHz bandwidth; the remaining spectral windows had 128 channels and 2~GHz bandwidth. J0423$-$0120 was used as the absolute flux and spectral bandpass calibrator. Phase switching to J0427+0457 was performed to correct the time-dependent phase variations. The total time on-target was 160~min, with approximately 40~min per epoch.

The data were calibrated using the ALMA pipeline within CASA. After confirming the quality of the pipeline calibration, the data were averaged in frequency by a factor of 16 to increase the signal-to-noise ratio per channel and the computational speed during imaging. This resulted in a spectral resolution of 22.6~km\,s$^{-1}$~channel$^{-1}$. Self-calibration was performed using the continuum spectral windows to spatially align the four epochs and correct for any residual time-dependent phase and amplitude errors. The continuum subtraction was performed by fitting a model to the continuum in the line-free spectral channels and subtracting it from the data in the visibility plane.

\begin{figure}
    \centering
    \includegraphics[width=0.48\textwidth]{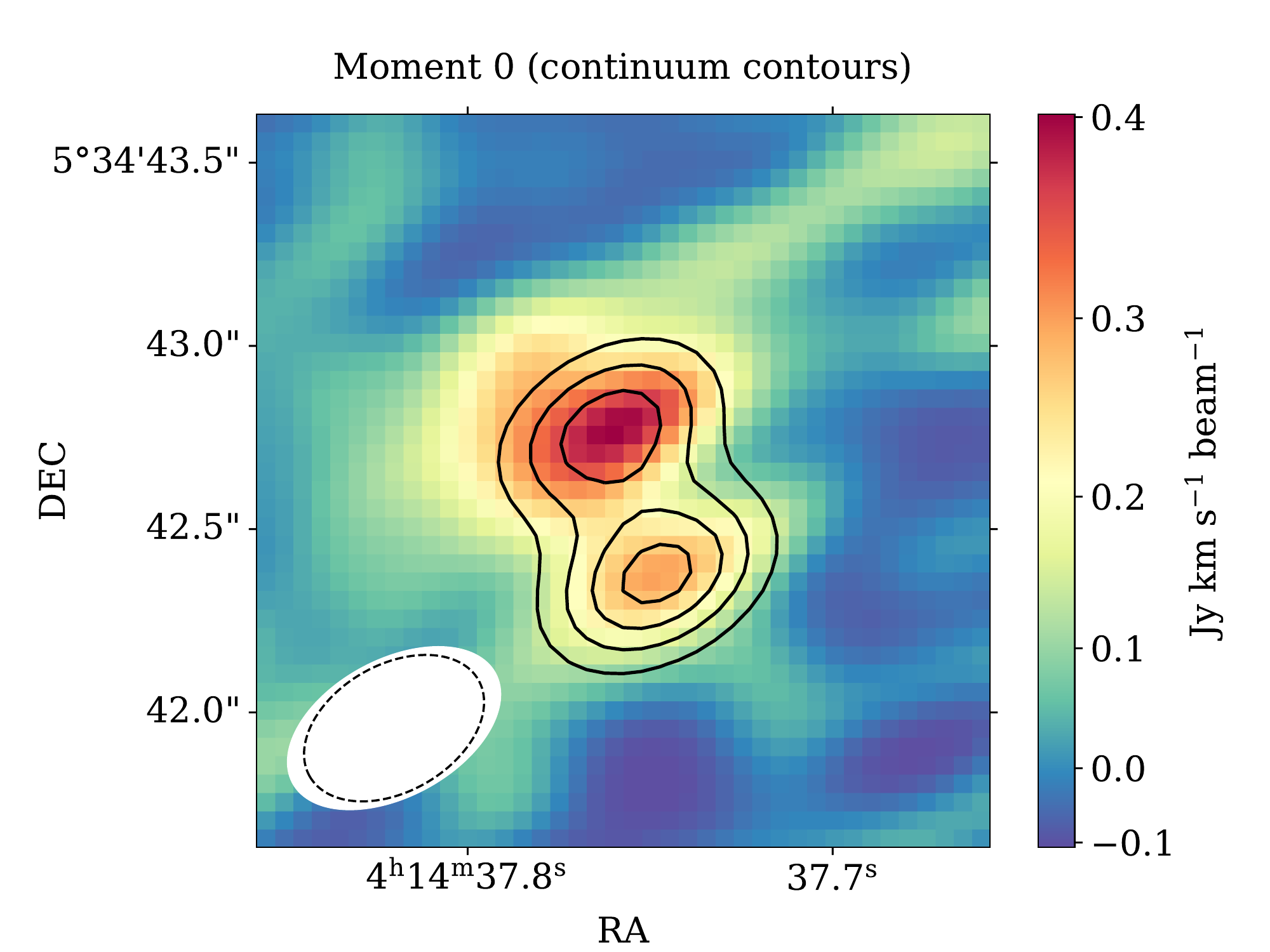}
    \includegraphics[width=0.48\textwidth]{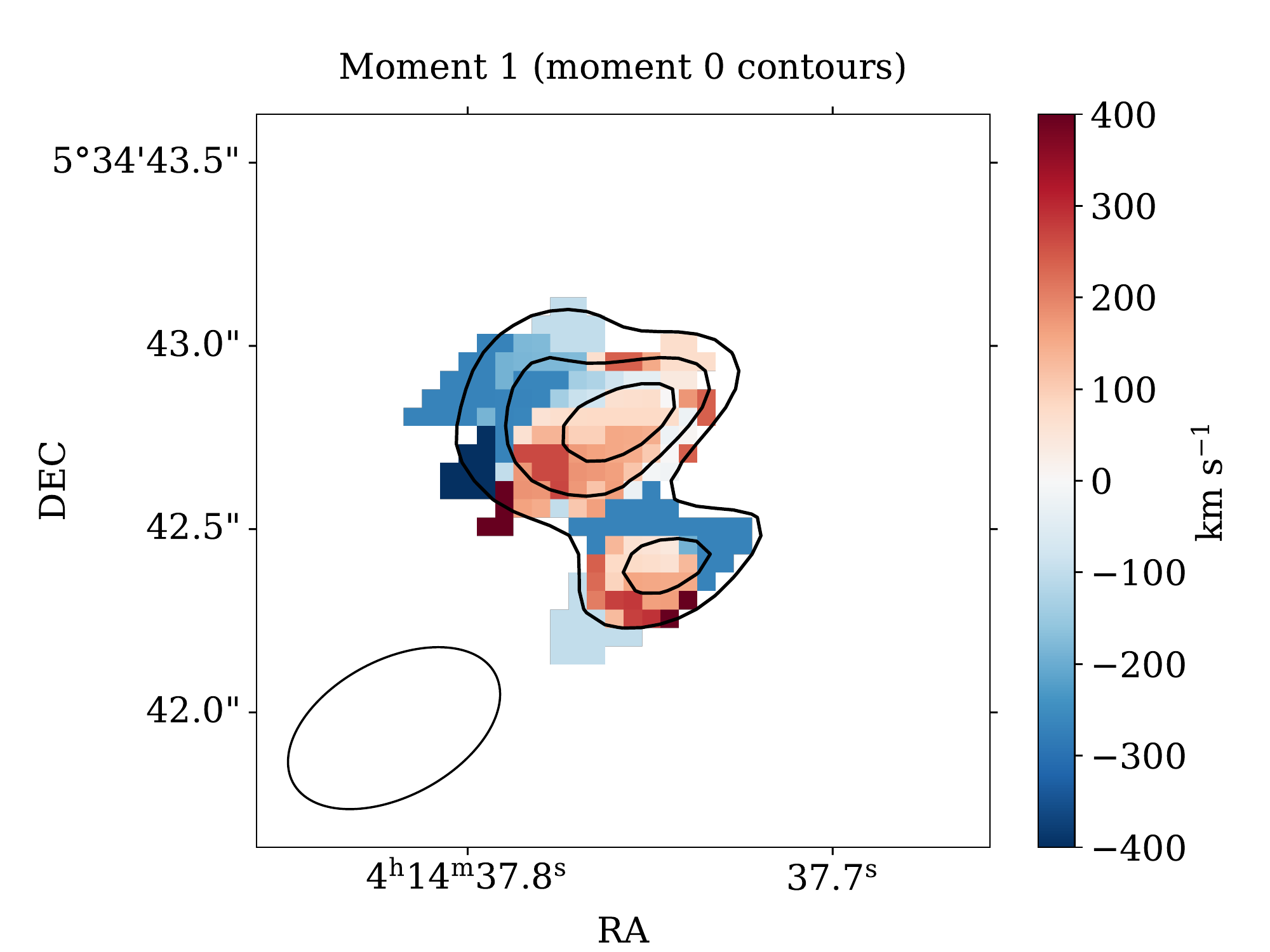}
    \includegraphics[width=0.48\textwidth]{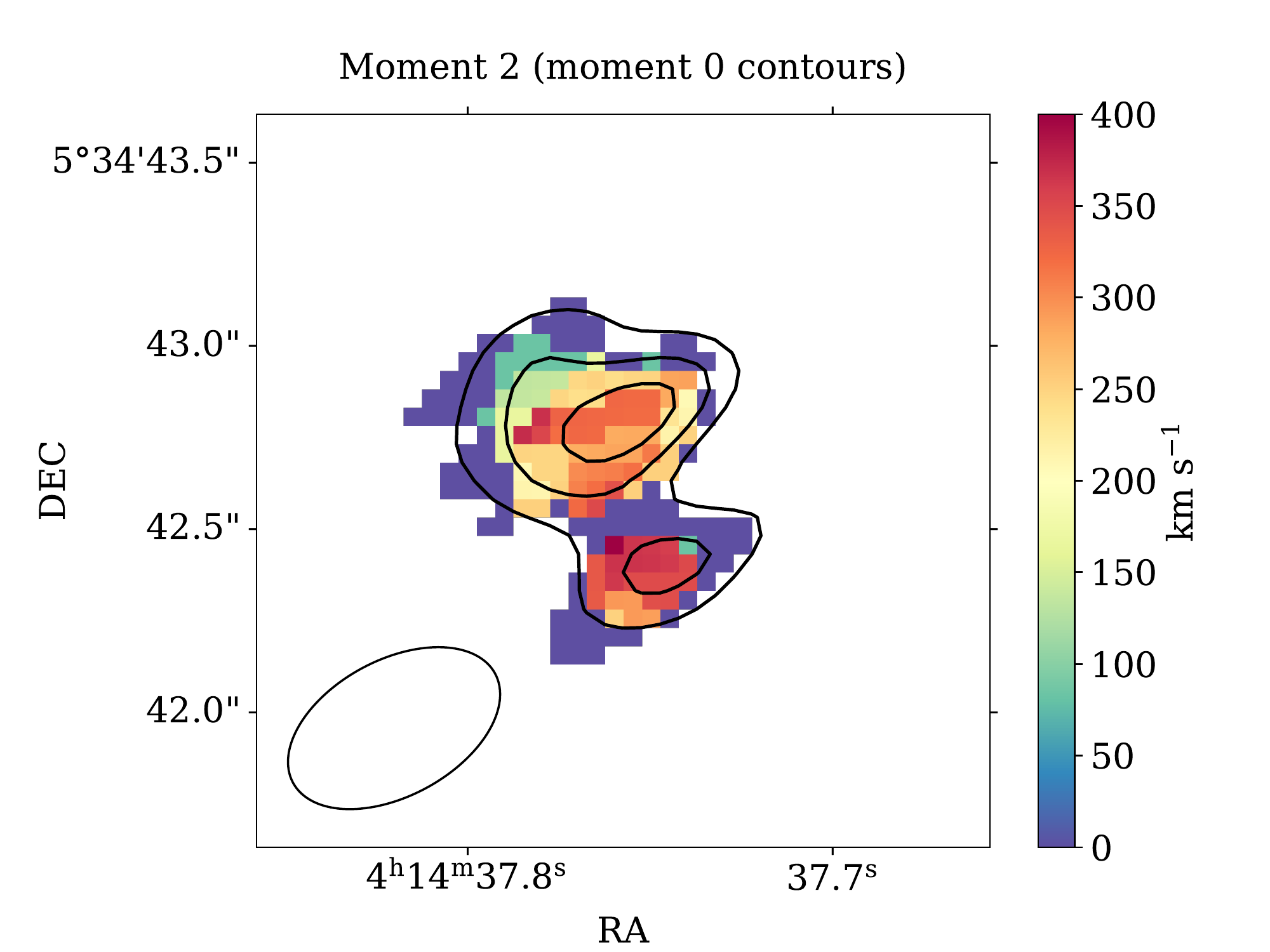}
    \caption{Velocity integrated line intensity (moment 0), velocity field (moment 1) and velocity dispersion (moment 2) of the H$_{2}$O~(4$_{14}$--3$_{21}$ line emission. The moment 1 and 2 images are created by masking emission below $3\sigma$, where $\sigma$ is the mean rms noise per channel. In the moment 0 image, contours showing the continuum are over-plotted with peak fractions 0.3, 0.5, 0.7, 0.9. The moment 1 and 2 images show contours of the moment 0 image at 5, 7, 9$\sigma$ (where $\sigma$ is the rms noise). The synthesised beam is shown in the lower left corner and has a size of 0.4~arcsec~$\times$~0.6~arcsec.}
    \label{fig:moments}
\end{figure}

The 100 GHz continuum image (not presented here) shows four point sources of emission, consistent with the findings of \cite{Kuo:2019} and with the far-infrared--radio spectral energy distribution that shows synchrotron-dominated emission at this frequency \citep{Stacey:2018a}. We find no significant variability in the continuum flux density between the four epochs. We do not detect the 5$_{23}$--6$_{16}$ H$_2$O line, consistent with the results of \citet{Kuo:2019}\footnote{This non-detection is not surprising as the excitation conditions ($E_{u}\sim3000$~K) mean that it is not likely to be detected from an extragalactic source.}. We confirm the detection of the 4$_{14}$--3$_{21}$ H$_2$O line; the line intensity and line luminosity (uncorrected for lensing magnification) are given in Table~\ref{table:lines}. We detect lensed images A1, A2 and B, but do not detect image C (the least magnified image). Moment maps of the line emission from lensed images A1 and A2, using a signal-to-noise ratio cut of 3$\sigma$, are shown in Fig.~\ref{fig:moments}.

\subsubsection{340~GHz data}

MG~J0414+0534 was observed with ALMA in Band 7 under project code 2013.1.01110.S (PI: Inoue). The spectral coverage included the CO~(11--10) emission line ($\nu_{\rm rest}=1267.014$~GHz). Details of the observations, data reduction and imaging are reported by \cite{Stacey:2018b} and are not presented here. The continuum imaging reveals four point sources of emission from the quasar and an Einstein ring of thermal dust emission. Imaging of the CO~(11--10) shows four compact components at the location of the quasar images. The resolved velocity field from the two merging images (A1 and A2) shows evidence of ordered rotation around $\pm500$~\kms\ from the systemic velocity. 

The positions of the redshifted and blueshifted components of the line emission were extracted by creating images from selected channels on either side of the systemic velocity. The positions of these components were extracted by fitting 2-dimensional Gaussians using the task {\sc IMFIT} within CASA.

\begin{table*}
\caption{Observed line intensities, line luminosities and continuum flux densities for the ALMA observations, uncorrected for lensing magnification. A 5~percent error in amplitude calibration should be assumed for the continuum flux density.}
    \begin{tabular}{ p{2cm} c c c c l }
        & $I_{\rm line}$ & $L_{\rm line}$ &$L'_{\rm line}$ & $S_{\rm cont}$ & Reference  \\
        & (Jy~km~s$^{-1}$) & (L$_{\odot}$) & (K~km~s$^{-1}$~pc$^{2}$) & (mJy) & \\ \hline
        H$_{2}$O~(4$_{14}$--3$_{21}$) & $1.0\pm0.1$ & $(5.3\pm0.6)\times10^{7}$ & $(3.0\pm0.4)\times10^{10}$ & 27 & this work \\
        H$_{2}$O~(6$_{16}$--5$_{23}$) & $0.30\pm0.03$ & $(9.5\pm0.9)\times10^{5}$ & $(2.7\times0.3)\times10^{12}$ & 71 & \cite{Castangia:2011} \\
        CO~(11--10) & $6.4\pm0.3$ & $(1.16\pm0.06)\times10^{9}$ & $(1.78\pm0.09)\times10^{10}$ & 24 & this work; \cite{Stacey:2018b} \\ \hline
    \end{tabular}
    \label{table:lines}
\end{table*}

\begin{figure}
    \includegraphics[width=0.48\textwidth]{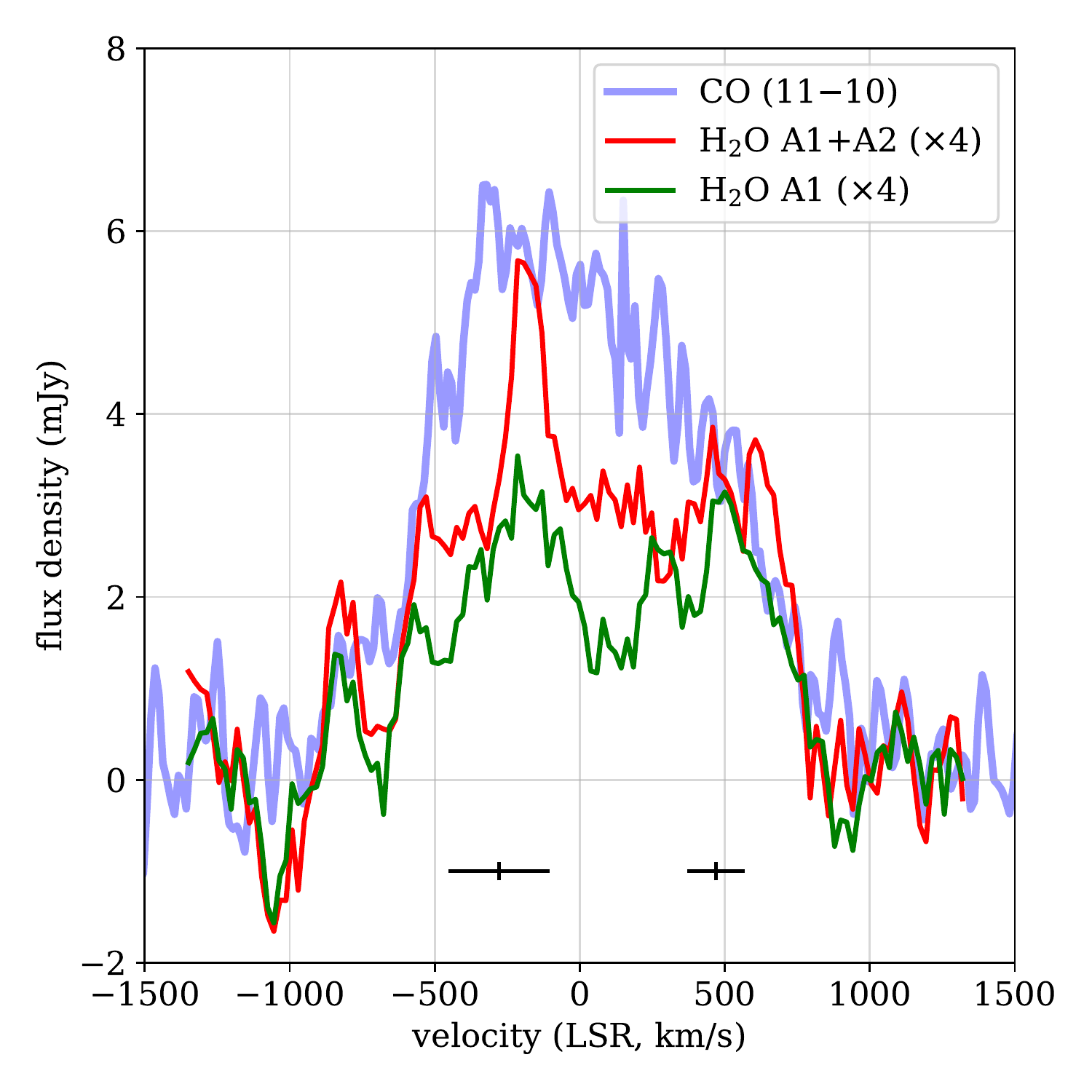}
    \caption{Line profiles of CO~(11--10) (blue) and H$_{2}$O~(4$_{14}$--3$_{21}$). The H$_{2}$O line integrated over both A1 and A2 is shown in red, and the line from the peak of A1 is shown in green. The black crosses indicate the peak and FWHM of the 22~GHz H$_2$O megamaser components from \protect\cite{Castangia:2011}. The H$_{2}$O line profiles are smoothed with a boxcar kernel of width 5 channels (113~km\,s$^{-1}$) and scaled by a factor of 4, for comparison. The systemic velocity is relative to $z=2.639$, using the radio definition of velocity.}
    \label{fig:line_profile}
\end{figure}

\begin{figure*}
    \centering
    \includegraphics[width=\textwidth]{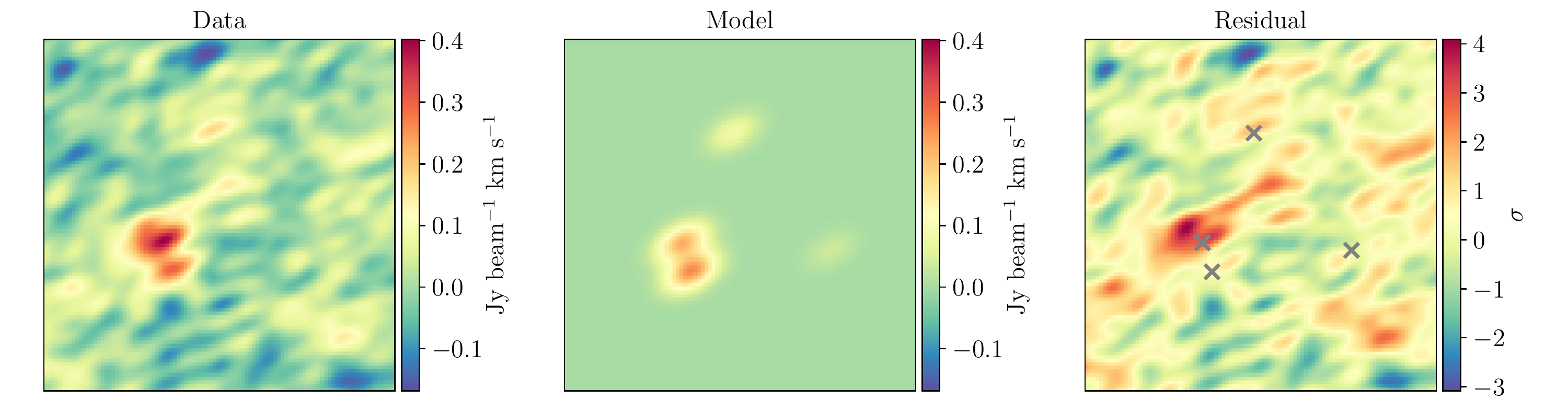}
    \caption{Gaussian model fit to imaging of the 380~GHz H$_2$O line integrated line intensity. {\it Left to right}: the data; the model on the same colour scale as the data; and the residuals (data$-$model) in units of $\sigma$, where $\sigma$ is the rms noise in the data. The Gaussian model components are fixed to the size of the synthesised beam (PSF) and their positions are fixed to the image positions in the continuum emission. The component flux densities are initially allowed to optimise, but afterwards are fixed to the continuum flux ratios reported by \citet{Stacey:2018b} relative to the optimised flux density of image A1. Residuals of this model fit show surface brightness features around image A2 at the $4\sigma$ level.}
    \label{fig:gauss_fit}
\end{figure*}

\begin{figure}
    \includegraphics[width=0.48\textwidth]{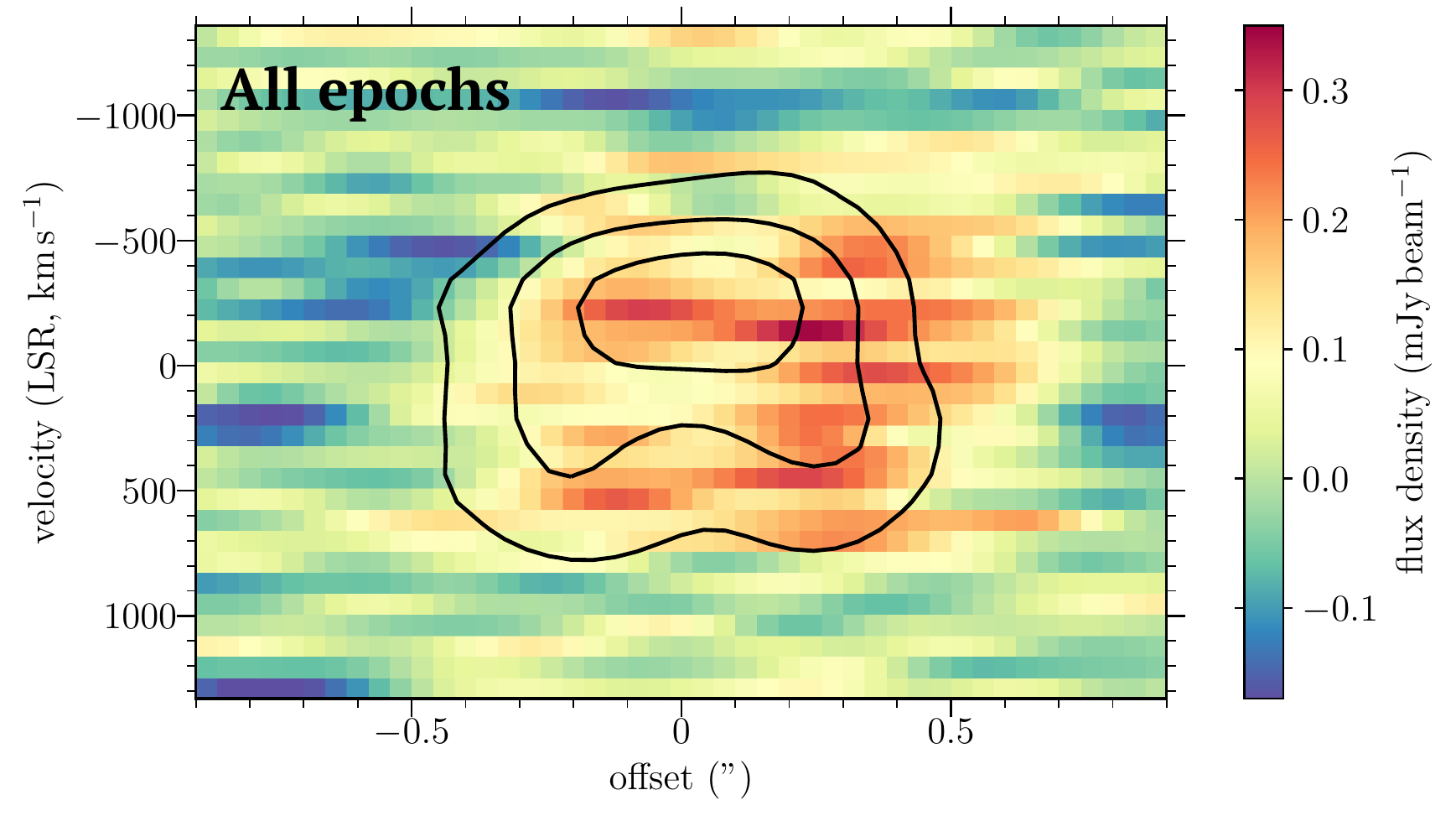}
    \includegraphics[width=0.48\textwidth]{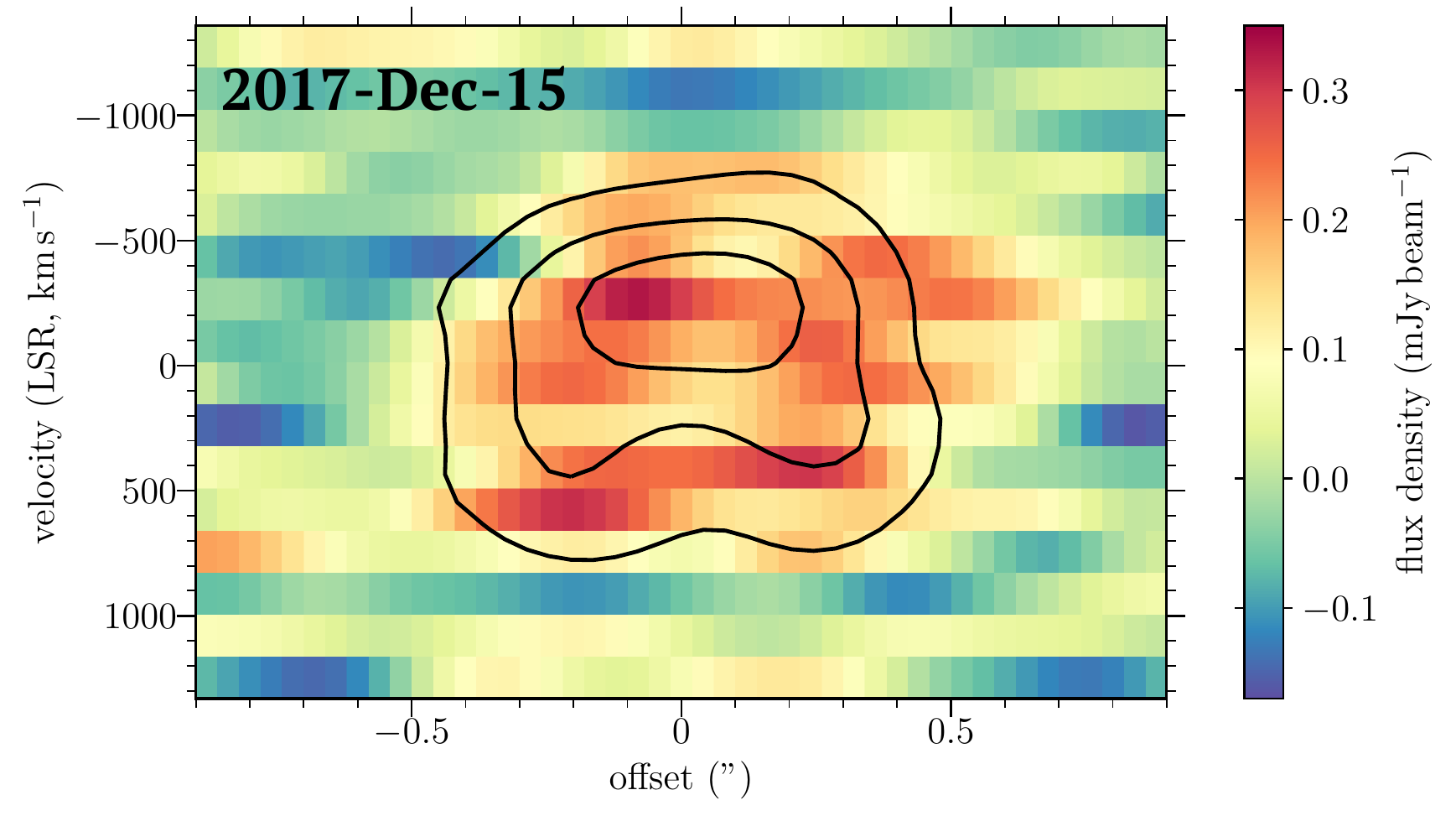}
    \includegraphics[width=0.48\textwidth]{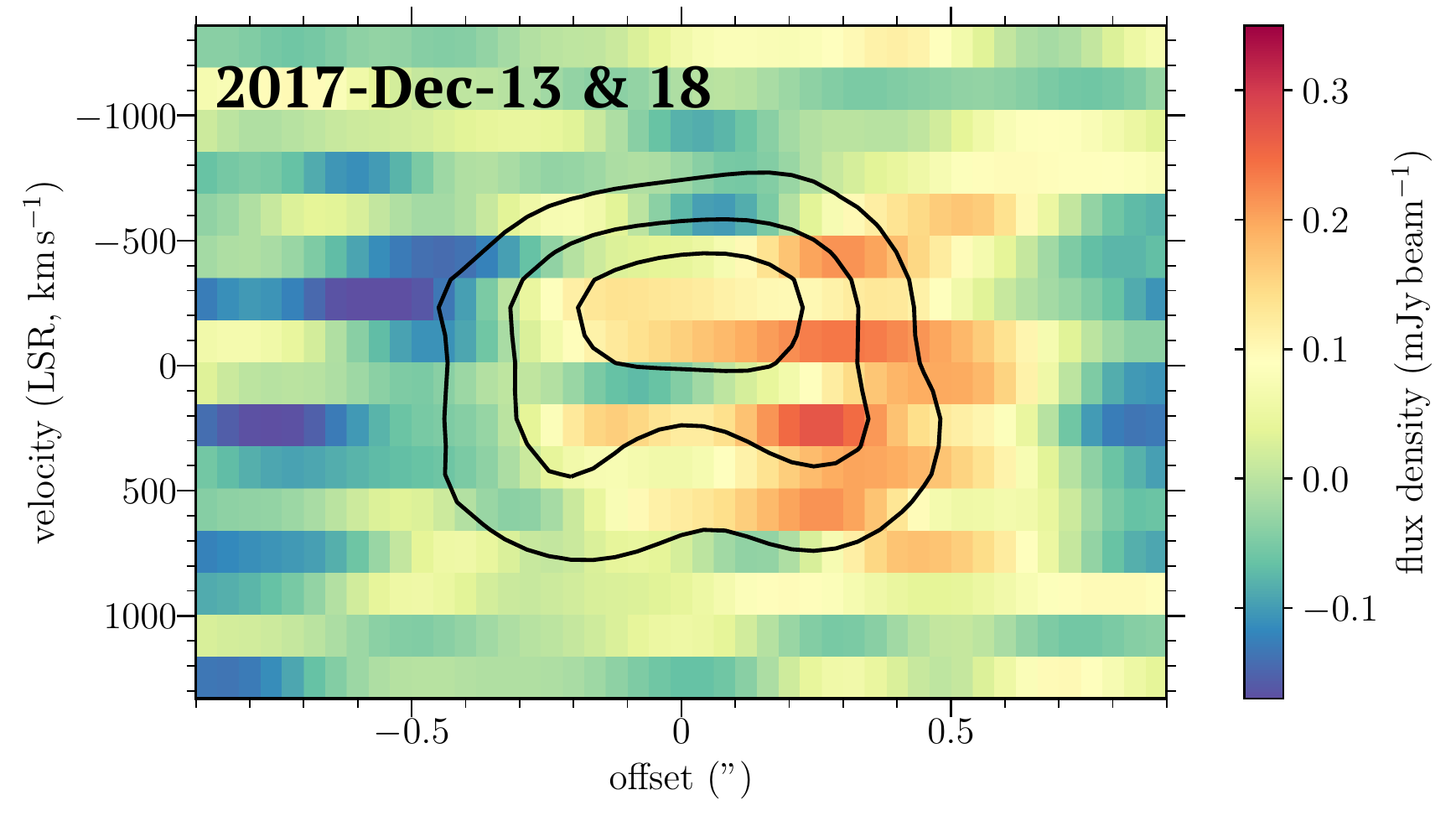}
    \caption{Position--velocity diagrams for the 380~GHz H$_{2}$O line emission from the merging images in the direction A1 to A2 (left to right). The top image shows the data from all 4 epochs; the middle image is data only from two epochs on 2017 December 15; the bottom image is data combined from 2017 December 13 and 18. The contours show the CO~(11--10) line emission from the same region, smoothed with a Gaussian kernel to match the angular resolution of the 380~GHz H$_{2}$O line emission.}
    \label{fig:pv_plot}
\end{figure}

\section{Results}
\label{section:results}

In this section, we present an analysis of the observed properties of the 380~GHz H$_2$O emission from MG~J0414+0534, and the inferred source-plane properties of the various molecular gas components, relative to the quasar and the radio jets.

\subsection{Velocity structure}

Fig.~\ref{fig:line_profile} shows the line profile of the 380~GHz H$_{2}$O~(4$_{14}$--3$_{21}$) emission, relative to the CO~(11--10) line profile from \cite{Stacey:2018b}, using an aperture including lensed images A1 and A2, and also from only lensed image A1 in the case of the former. The 380 GHz H$_{2}$O line profile obtained from only image A1 has a structure suggestive of a double horn profile, characteristic of emission from a disc. We fit two Gaussian components to the double-horn profile isolated from image A1, which are centred on $-210\pm20$~\kms\ and $490\pm20$~\kms, with full width at half maximum (FWHM) of $630\pm50$~\kms\ and $420\pm40$~\kms, respectively. The CO line profile from both A1 and A2 is more peaked around the systemic velocity compared to the total 380 GHz H$_{2}$O emission, without a double-horn shape. The overall FWHM of the 380~GHz H$_{2}$O line profile is $\sim1000$~\kms, comparable to that observed from the CO~(11--10) line (i.e. $1100\pm100$~\kms, based on a Gaussian fit). The similar terminal velocities suggest that they are associated with the same region of the source.

The double-peaked 380~GHz H$_{2}$O profile from image A1, under the assumption that the emission is from a disc, suggests a correction to the systemic velocity of $+140\pm10$~\kms. The centre of the CO~(11--10) profile also suggests a slightly redshifted systemic velocity. We note that the peaks in the 380~GHz H$_{2}$O line profile are consistent with the blue and redshifted components of the 22~GHz H$_2$O megamaser system reported by \cite{Castangia:2011} (shown with black crosses in Fig.~\ref{fig:line_profile}), but with larger associated dispersions.

From the moment maps of the 380~GHz H$_2$O line emission (see Fig.~\ref{fig:moments}), we find that the line emission is only marginally resolved in total intensity for the two merging lensed images (which are also those with the highest magnification; see below). However, there is evidence of resolved structure in the velocity map, where we find a (red) component at around +100~\kms, that is, close to the systemic velocity, which is coincident with the peak in the line total intensity, and a spatially-offset (blue) component that is at a velocity of around $-300$~\kms. This velocity structure is markedly different from the CO (11--10), where there is evidence of only a disc component associated with the AGN. This could be, in part, due to the better angular resolution of the CO (11--10) observations, where the overall velocity structure is less beam-smeared. We see that the velocity dispersion of the 380~GHz H$_2$O emission peaks at around $400$~\kms\ and is co-spatial with the quasar, both of which are consistent with the properties of the CO (11--10) emission. This further suggests that the major component of these two molecular gas tracers are likely associated.

While the angular resolution of the 380~GHz H$_2$O line observations are around a factor of 2 lower than those of the CO~(11--10), the water emission cannot be significantly more extended (as it would form an arc or ring), or be spatially offset (which would produce a different image configuration). Assuming a smooth distribution of mass in the lens, these merging images should be almost mirror images with a similar flux density. Consistent with \cite{Kuo:2019}, the H$_2$O line emission from image A2 appears around twice as bright as A1: this flux density ratio is reversed compared to the radio/mm continuum, CO~(11--10) emission and multi-wavelength archival observations \citep{Stacey:2018b}. As the 380~GHz H$_{2}$O emission is spatially coincident with the CO~(11--10) and mm-continuum emission from the quasar, small-scale mass structure in the lens is unlikely to be responsible for the different flux ratio as the emission is subject to the same local lens potential. Therefore, the observed reversed flux ratio is more likely due to the effect of source structure rather than the mass distribution in the lens (see below).

We fit Gaussian components to the four lensed images in the 380~GHz H$_2$O line integrated intensity to determine at what level the emission is spatially resolved. Assuming the 380~GHz H$_2$O line emission that is coincident with the quasar images is unresolved and has image flux ratios consistent with the mm-continuum, we fix four Gaussian components to the size and position angle of the synthesised beam, and fix the component positions to the centroids of the mm-continuum emission. We initially allow the flux density of these components to optimise. We then fix the flux density of images A2, B and C using the flux ratios reported by \cite{Stacey:2018b}, normalised relative to the optimised flux density of A1. As shown in Fig.~\ref{fig:gauss_fit}, the residuals of this model fit have surface brightness features at the 4$\sigma$ level, suggesting that some 380~GHz H$_2$O line emission is spatially resolved around image A2.

Fig.~\ref{fig:pv_plot} shows position-velocity diagrams of a slice through images A1 to A2 for the 380~GHz H$_2$O line emission, with the contours of the CO~(11--10) emission from the same region overlaid. A double-peak structure can be seen within the extent of the CO disc, as suggested by the line profile. The CO emission has a caret shape, due to the fact that the blueshifted emission is closer to the caustic and is more strongly magnified. An additional velocity component from the 380~GHz H$_2$O emission can be seen around image A2 that is more extended than the CO emission and does not have a mirror component in A1. This is the extended emission seen in the moment 0 image and contributes the narrow peak in the line profile at around $-220$~\kms\ (see Fig.~\ref{fig:line_profile}). The anomalous velocity component is not consistent with the caret shape of the CO~(11--10) disc and is spatially offset. Therefore, this cannot be emission from the disc.

\subsection{Variability}

\begin{figure*}
    \centering
    \includegraphics[width=0.49\textwidth]{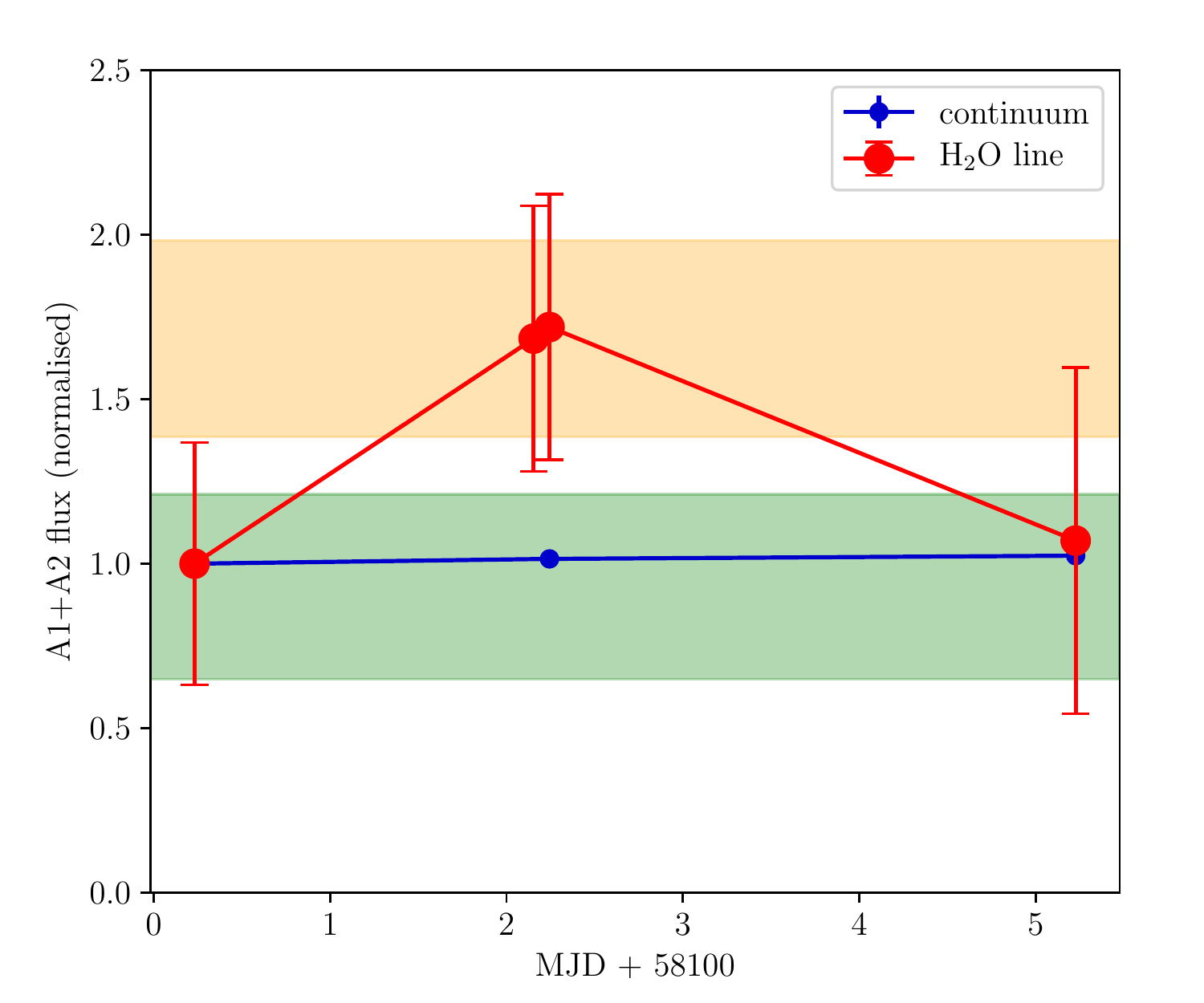}
    \includegraphics[width=0.49\textwidth]{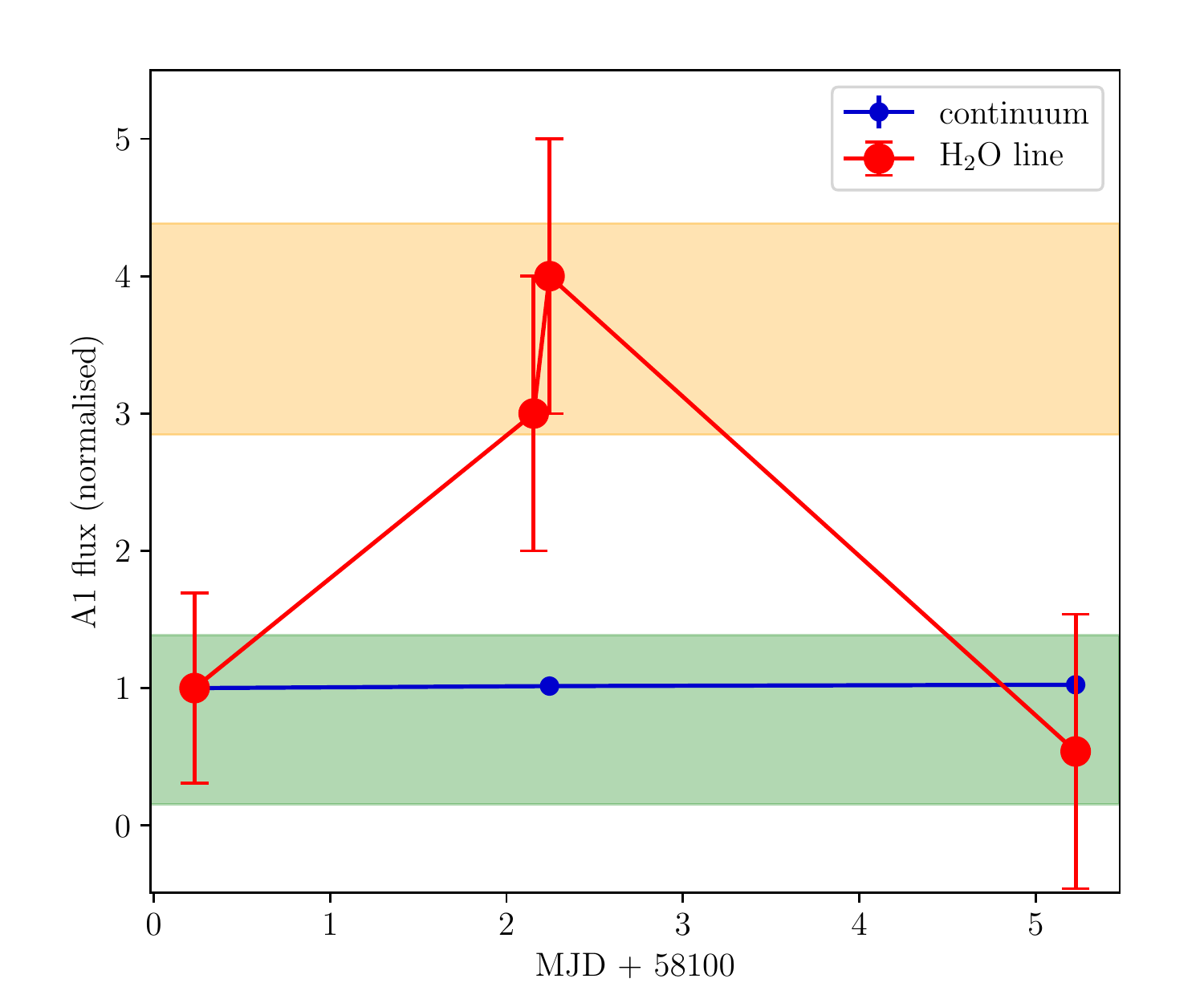}
    \caption{The normalised, velocity-integrated flux density of the 380~GHz H$_2$O line emission and rest-frame continuum for the four epochs. We show the integrated flux density of both merging images (A1+A2; left) and for image A1 only (right). The 380~GHz H$_{2}$O line emission is shown in red and the continuum is shown in blue. The orange region shows the combined flux density and uncertainty for both epochs on 2017 December 15; the green region shows the combined flux and uncertainty for 2017 December 13 and 18. }
    \label{fig:variability}
\end{figure*}

As the observations of the 380~GHz H$_2$O line were obtained in four epochs with separations between a few hours and 6 days, we examine the individual observations to look for short time-scale variability. The emission from the merging images (A1+A2) appears to increase by a factor $1.7\pm0.8$ between the observations taken on 2017 December 13 and 15, but this is not significant as the uncertainty is large. We find tentative evidence for variability when comparing the integrated flux from combining the two 2017 December 15 epochs with the combined preceding and succeeding epochs on 2017 December 13 and 18 (combined to match the signal-to-noise of 2017 December 15 epochs). The significance of this variability is at the $2\sigma$ level when we consider the total integrated emission (see Fig.~\ref{fig:variability}, left panel). However, the significance rises to $3\sigma$ when we only consider the integrated flux from image A1 (Fig.~\ref{fig:variability}, right panel). We do not find evidence that the continuum emission is variable over the observations, which is consistent with the results obtained by \citet{Stacey:2018b} for the rest-frame 1267~GHz continuum emission from the AGN. We note that, although differences in amplitude scaling could mimic source variability, this would not affect the signal-to-noise ratio of the line emission.

The strongest evidence for variability in the 380~GHz H$_2$O line is seen in the position-velocity diagrams for these observations, which are also shown in Fig.~\ref{fig:pv_plot}. The emission can be seen to be quite different between the data sub-sets. The emission in the combined data from the epochs on 2017 December 15 is higher overall, but, in particular, the emission from image A1 shows the largest difference. The time delay between the merging images of the quasar is expected to be 0.4~days, based on our lens model (see below). When comparing the two epochs taken on 2017 December 15, the flux ratio appears to switch, with A2 brighter in the first epoch. This accords with our expectation from the lens model (image A2 is the leading image), but these measurements are within the overall uncertainties. Our findings could be interpreted as two components of 380~GHz H$_2$O line emission, one that is variable on timescales of less than a few days, as seen in image A1, and one that is constant over the timescale of the observations.

\subsection{Source-plane structure}

In order to determine the structure of the various emission components from MG~J0414+0534 in the source plane, we first determine a model for the lens. This was performed using the parametric lens modelling software {\sc gravlens} \citep{Keeton:2001a,Keeton:2001b}. We assume a simple model of a singular isothermal ellipsoid with external shear (SIE+$\gamma$) for the primary lens (G1), and a singular isothermal sphere (SIS) for the satellite galaxy (G2). The 1.6~GHz VLBI positions, having the highest precision and largest number of image components, were used to optimise for the best lens model parameters. The flux densities of the components were not used in the modelling, as these are known to deviate from a smooth lens model \cite[e.g.][]{Macleod:2013}. While an improved fit to the image positions may also be obtained with more complex parametric models, such as an ellipsoidal power-law mass distribution \citep[e.g.][]{Spingola:2018}, this will not significantly change our interpretation here. The resulting best fit parameters are given in Table~\ref{table:lens_params}, and are found to be in good agreement with previous modelling of the system (e.g. \citealt{Ros:2000,Macleod:2013}). As the absolute astrometry between the various data sets is lost after self-calibration, we align them using the compact core emission that is associated with synchrotron emission from the quasar. The consistent image separation of these components in each data set supports the assumption that the component we identify as the radio core is coincident with the quasar.

\begin{figure*}
    \centering
    \includegraphics[width=\textwidth]{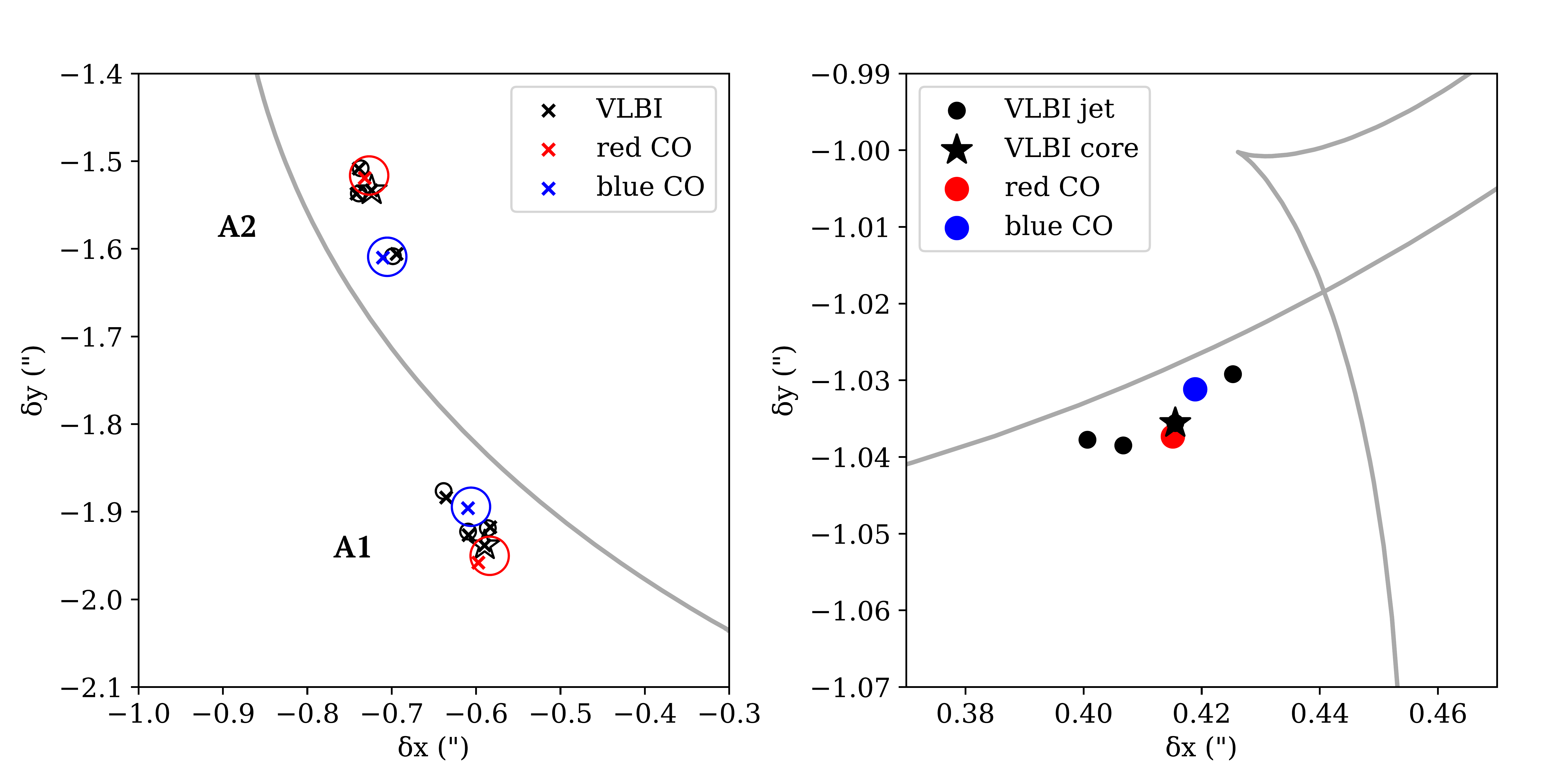}
    \caption{Lens-plane (left) and source-plane (right) positions of the CO~(11--10) and 1.6 GHz VLBI components. We show only the region of the lens plane around the merging images (A1 and A2). The grey lines trace the critical curve on the lens plane and tangential caustics on the source plane. The observed image positions are shown with crosses. The model-predicted positions on the lens and source plane are shown in red and blue circles for the respective CO~(11--10) velocity components, and in black circles for the 1.6~GHz VLBI components. The source plane structure suggests a molecular gas disc of CO~(11--10) emission bisected by two perpendicular radio jets.}
    \label{fig:vlbi_model}
    \centering
    \includegraphics[width=\textwidth]{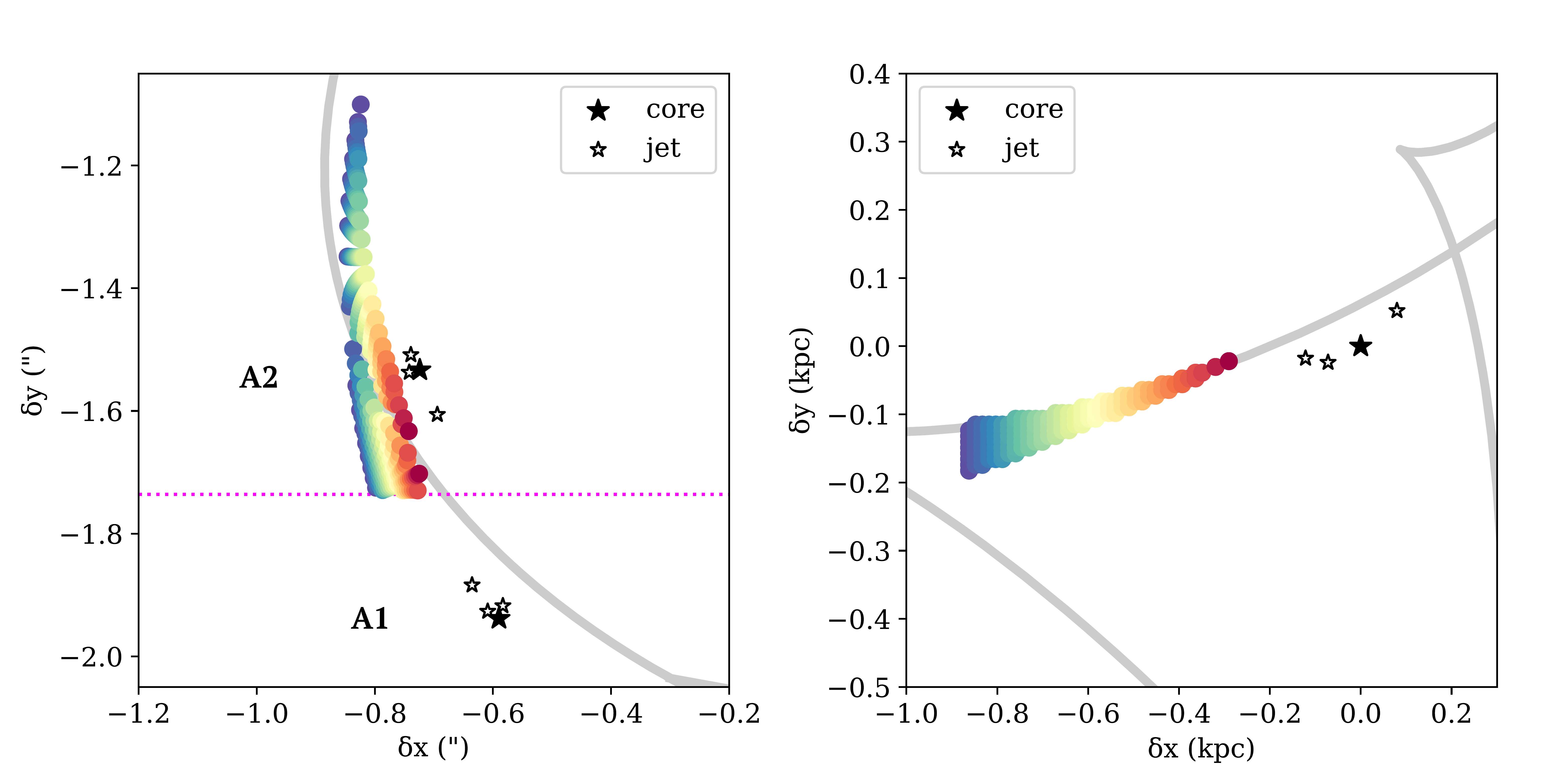}
    \caption{Lens-plane images (left) corresponding to mock positions of the anomalous 380~GHz H$_2$O velocity component on the source plane (right). The radio core and jets are shown with filled and empty stars, respectively, and the grey lines trace the critical curve and tangential caustics. The source-plane positions are shown in kpc relative to the radio core. Sources that produce images below the midpoint of A1 and A2 (shown with the dotted magenta line) are excluded. We also exclude sources that produce fewer than four images, as they would not produce an image on the relevant area of the lens plane or would be a very large distance from the quasar.}
    \label{fig:steps}
\end{figure*}

The reconstructed source-plane emission from the 1.6~GHz VLBI continuum and CO (11--10) velocity components are shown in Fig.~\ref{fig:vlbi_model}. We find that the jet and counter jet (the components either side of the radio core) form an almost-linear structure that is extended in an East--West direction by around 200~pc in projection. The red and blue CO (11--10) velocity components are almost perpendicular to the radio jets, and form what we interpret as a molecular gas disc with a size that is 60~pc in projection around the quasar. Given the similarity in the position-velocity diagrams of the CO (11--10) and the 380 GHz H$_2$O emission, we assume that the molecular gas disc also contains part of the water emission observed from MG~J0414+0534.

Assuming just half of the 380 GHz H$_2$O emission we detect originates in a disc, and assuming the total magnification of the radio core inferred from our lens modelling is representative of the magnification of the H$_2$O disc ($\mu_{\rm tot} \simeq 40$), we estimate the isotropic line luminosity is $\sim7\times10^{5}$~L$_{\odot}$.

\begin{table}
    \caption{The best parameters of the lens model, based on the positions of the VLBI image sub-components. The primary lens, G1, is modelled with an SIE and external shear, and the secondary lens, G2, is modelled with an SIS. $b$ is the Einstein radius of the lens. The positions ($x$,$y$) are given relative to the brightest component of image B. $e$ and $\gamma$ are the ellipticity and external shear, and $\theta_{e}$ and $\theta_{\gamma}$ are their respective position angles East of North.}
\adjustbox{width=0.46\textwidth}{
    \begin{tabular}{l c c c c c c c c c c}
        & $b$ & $x$ & $y$ & $e$ & $\theta_{e}$ & $\gamma$ & $\theta_{\gamma}$ \\
        & (") & (") & (") & & ($^{\circ}$) & & ($^{\circ}$) \\ \hline
        G1 & 1.076 & $+$0.482 & $-$1.279 & 0.235 & $-$84.16 & 0.098 & 53.85 \\
        G2 & 0.196 & $+$0.913 & $+$0.150 & - & - & - & - \\ \hline
    \end{tabular}}
    \label{table:lens_params}
\end{table}

To constrain the origin of the anomalous velocity component in the 380~GHz H$_2$O line emission (i.e. emission not from the disc) we place a grid of mock sources on the source plane and compute the positions predicted by our lens model. Fig.~\ref{fig:steps} shows the image positions produced by the mock sources relative to the radio components. As the anomalous velocity feature is not detected from the emission peak of A1, we exclude all source positions that produce an image below the midpoint between image A1 and A2. We also exclude any positions that are doubly imaged (outside the tangential caustic), as these would not produce an image in the region of interest. Doubly imaged source positions that could produce an image near the merging images would be on the opposite side of the caustics (several kpc from the quasar, as suggested by \citealt{Kuo:2019}), which we do not consider to be feasible. We find that the remaining viable source positions are very close to the lens caustic, which would produce a high magnification of the anomalous velocity component. However, as the position-velocity diagrams (see Fig.~\ref{fig:pv_plot}) and Gaussian source fitting to the images (see Fig.~\ref{fig:gauss_fit}) show the anomalous feature is North of A2 and around 0.3--0.4~arcsec from the midpoint, only the green and blue source regions are in agreement with the observations. This suggests that the anomalous 380~GHz H$_2$O feature originates $>600$~pc East of the quasar and perpendicular to the molecular gas disc. 

Assuming that half of the 380~GHz H$_2$O emission originates in the anomalous feature, we take the highest combined magnification of the merging images produced by the mock sources (about 330) to infer a minimum intrinsic isotropic line luminosity of $8\times10^{4}$~L$_{\odot}$.

\subsection{Size and structure of the molecular gas disc}

From our lens modelling, we find that the CO~(11--10) originates in a molecular gas disc with a size (diameter) of 60~pc around the quasar. From this, we estimate the enclosed dynamical mass ($M_{\rm dyn}$) of the system using
\begin{equation}
    M_{\rm dyn} = \frac{R V_{\rm max}^{2}}{G} ,
\label{equation:mdyn}    
\end{equation} where $R$ is the 30~pc radius (60~pc diameter) and $V_{\rm max}$ is the maximum rotational velocity. We assume the rotational velocity is the line velocity at the FWHM ($V_{\rm obs}=550$~\kms), corrected with an inclination angle of 45~deg (i.e. $V_{\rm max}=V_{\rm obs} /  \sin45^{\circ}$). We find a dynamical mass of $M_{\rm dyn}(<30~{\rm pc}) \sim3\times10^{9}$~M$_{\odot}$, which is only three times larger than the black hole mass determined by \cite{Pooley:2007} from a microlensing analysis of the X-ray emission from the quasar accretion disc. This suggests that the kinematics of the CO~(11--10) emission are driven by the properties of the central supermassive black hole. 

The resolved velocity structure of the CO~(11--10) line, presented by \cite{Stacey:2018b}, shows a radially increasing velocity profile and appears to show a very high central velocity dispersion. However, 2-dimensional velocity fields are strongly affected by beam-smearing effects that can lead to an incorrect interpretation of the global dynamical properties \citep*{Lelli:2010}. Therefore, we defer a more comprehensive investigation of the disc dynamics to a future work that will use higher resolution imaging of the CO (11--10) line.

Assuming the molecular gas disc also contains water emitting regions with a Keplerian rotation and whose enclosed mass is dominated by the black hole (e.g. \citealt{Kuo:2011}), we use the same formalism as Eq.~\ref{equation:mdyn} to estimate the minimum size of the disc. Here, we assume the maximum rotational velocity is half the separation of the line peaks identified in Section~\ref{section:results}, corrected for an assumed disc inclination angle of 45~deg. This implies the size (diameter) of the water emitting disc is around 35~pc, that is, about half the size of the CO~(11--10) disc. Further high-resolution imaging of the 380~GHz H$_2$O emission with ALMA could confirm that the double peak structure is indeed part of a disc, as implied by the position-velocity diagrams shown in Fig.~\ref{fig:pv_plot}.

\section{Discussion and conclusion}
\label{section:discussion}

We have re-analysed spectral line imaging with ALMA of the rest-frame 380~GHz H$_2$O emission from MG~J0414+0534 ($z=2.639$), a gravitationally lensed dusty star-forming galaxy that hosts a type~1 quasar. These observations were first presented by \cite{Kuo:2019}, who reported a tentative detection of the line. \citeauthor{Kuo:2019} found that the broad line profile made the continuum subtraction challenging, and that the flux density ratio of the line emission in the two merging lensed images was reversed relative to multi-wavelength continuum and CO (11--10) observations for this object. 

\subsection{Origin of the H$_2$O emission}

From a careful re-analysis of these data, we find strong evidence that the 380~GHz H$_2$O line detection is robust. We independently verified the calibration and imaging (of both the target and the calibrators), and compared the line profile and position-velocity diagram with those from the CO~(11--10) and the 22~GHz H$_2$O megamaser that were previously reported for this object \citep{Stacey:2018b,Castangia:2011}. We find that the line width ($\sim1000$~\kms) and velocity dispersion ($\sim400$~\kms) are almost identical to the CO~(11--10) emission, and there are peaks in the line profile (at $-210$ and $+420$~\kms) that correspond to the velocities of 22~GHz H$_2$O megamaser components. Such similarities provide circumstantial evidence that the detection is robust. However, the strongest evidence comes from the position-velocity diagram, where there is evidence of a resolved velocity structure that is consistent with a rotating disc. This disc has an inferred size of around 35~pc, given the mass of the black hole, which is again similar to the resolved CO~(11--10) emission. We also find evidence for an additional 380~GHz H$_2$O component of emission, which, with lens modelling, is estimated to be $\sim600$~pc from the quasar. The source of the offset feature is unclear; perhaps corresponding to a jet-cloud interaction or a dense region of star formation. Its location close to the lens caustic suggests it could be very highly magnified by a factor of $>100$. 

Our findings may imply that the host galaxy of MG~J0414+0534 contains a significant number of water emitting regions, and our detection of this particular component is due to a chance lensing configuration. A similar scenario was suggested by \cite{Kuo:2019} in their analysis of the same data set, however, they proposed three water-emitting regions separated by a larger distance of around 1.5~kpc that are doubly-imaged. While the model we propose suggests a chance high magnification for the anomalous component, the doubly-imaged regions in the model of \citeauthor{Kuo:2019} produce much lower magnifications corresponding to higher intrinsic luminosities. Future observations of 380~GHz line emission from dusty star-forming galaxies will determine the rarity of high-luminosity water-emitting regions and test which scenario is more probable.

\subsection{Structure of the circumnuclear gas disc}

\begin{figure*}
    \centering
    \includegraphics[width=0.8\textwidth]{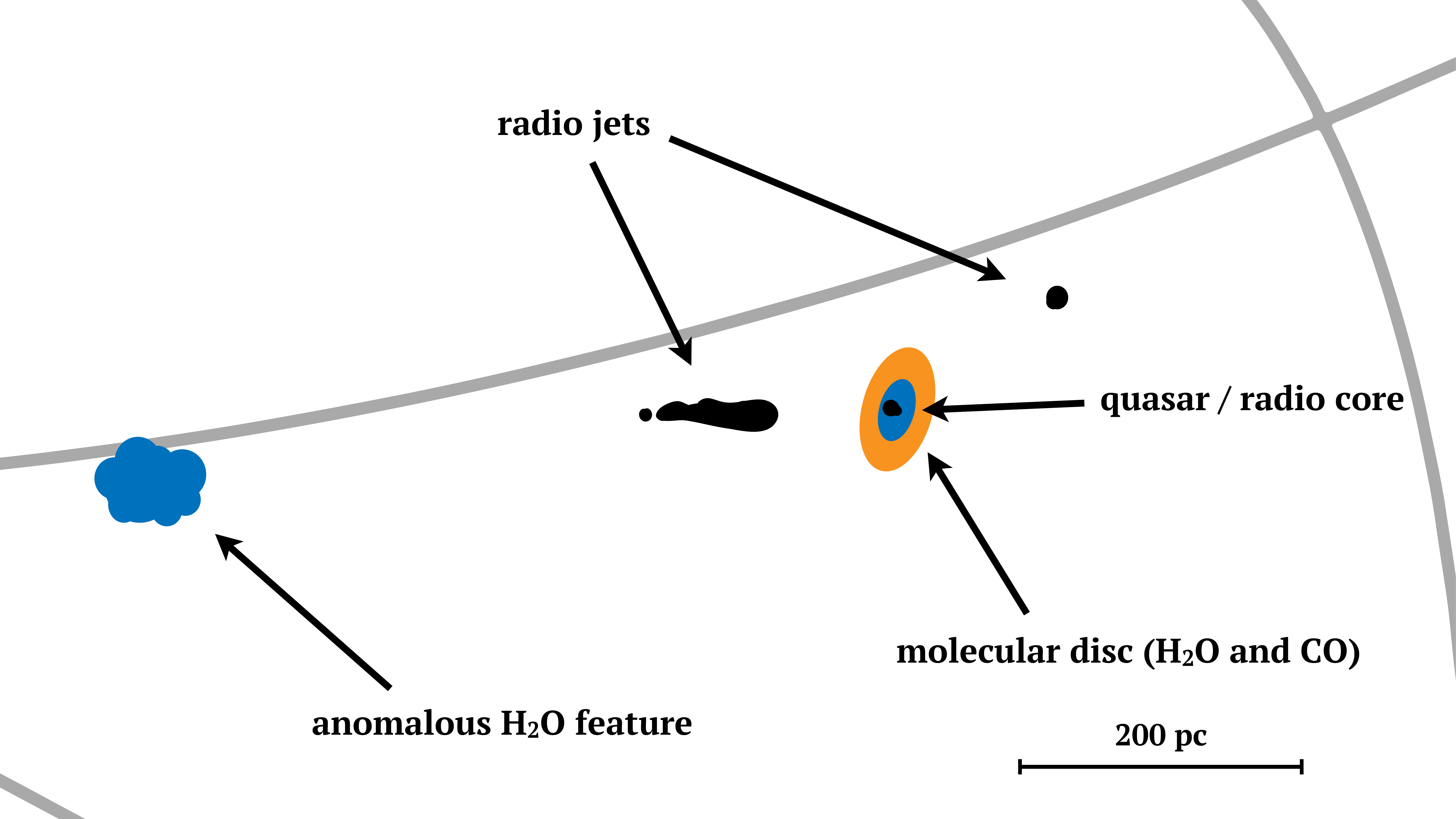}
    \caption{A toy model (approximately to scale) to explain the observed H$_2$O (blue) and CO~(11--10) (orange) line emission, relative to the radio core and jets (black) from MG~J0414+0534. This model predicts a molecular gas disc around the AGN with a size of 60~pc, with the 22 and 380~GHz H$_2$O emission enclosed in a disc of around 35~pc diameter. An additional component of 380~GHz H$_2$O emission is predicted to form around 600~pc in projection from the quasar, and may be associated with a highly magnified star-forming region that happens to lie close to the lens caustics (grey lines). The radio jets extend out to 200~pc in projection from the radio core, and are not expected to be associated with the highly magnified region of water emission from the system.}
    \label{fig:toy}
    \end{figure*}

We extend the investigation of the 380~GHz H$_2$O line emission in MG~J0414+0534 by combining these data with the information provided by the CO~(11--10) spectral line emission and global VLBI imaging of radio jet components to construct a model for the structure of the lensed quasar. Fig.~\ref{fig:toy} shows a toy model of the reconstructed source emission (approximately to scale). This model consists of a molecular gas disc around the quasar with 22~GHz and 380~GHz H$_2$O line emission with a size of around 35~pc, with the CO~(11--10) gas emission extending to a diameter of 60~pc. Bi-polar compact radio jets extend from the quasar radio core to a distance of about 200~pc, almost perpendicular to the gas disc. The other source of 380~GHz H$_2$O line emission is offset by $\sim600$~pc East of the quasar. 



We find tentative evidence for variability in the 380~GHz H$_2$O disc at the 3$\sigma$ level on timescales of hours to days. Amplitude fluctuations were previously reported for the 22~GHz H$_2$O megamaser in long-term monitoring \citep{Castangia:2011}, although the 1~month cadence could mask shorter-term variability. A high time variability is inconsistent with our estimate of $>35$~pc for the diameter of the 380~GHz H$_2$O disc and suggests a much more compact region with a size of just a few light days. Such high time variability has been detected from compact maser discs in the local Universe (e.g. \citealt{Braatz:2003}). Monitoring with a short-term cadence, such as with single dish observations, could determine whether the tentative variability is real.

The compact size of the molecular gas disc and the excitation conditions of the CO~(11--10) and water lines ($n_{\rm H_{2}}\sim10^{6-8}$~cm$^{-3}$; $E_{\rm u}\sim300$--600~K) suggest that we are probing a dense and warm environment close the AGN. Around one third of the dynamical mass enclosed by the CO disc is in the central supermassive black hole, perhaps indicating that we resolve a circumnuclear disc or the outer AGN torus. High densities close to AGN are often associated with maser activity in the local Universe. However, we find no clear evidence that the 380~GHz H$_2$O line observed here originates from a population of masers: the extremely high integrated luminosity is more consistent with observations of thermally excited H$_{2}$O lines from dusty star-forming galaxies (e.g. \citealt{Jarugula:2019}). The 380~GHz H$_2$O line is predicted to produce both thermal and non-thermal maser emission, and exhibits a sharp transition between these processes with density \citep{Gray:2016}. Further investigations of the 380~GHz H$_2$O line are needed to determine whether it is expected to be associated with thermal emission from dusty star-forming galaxies at high redshift, rather than non-thermal maser emission (or a combination of both).

Circumnuclear discs of dust and molecular gas have been resolved on pc-scales around AGN in the local Universe, which provide insight into the nuclear structure of AGN and the mechanisms of black hole feeding \citep{Gallimore:2016,Imanishi:2018,Garcia-Burillo:2016,Garcia-Burillo:2019,Combes:2019,Impellizzeri:2019}. A detection of a compact disc around the quasar of MG~J0414+0534 could suggest that the 380~GHz H$_2$O line is a useful tool to investigate the nuclear structure of AGN at high redshift. However, the low surface brightness of this 380~GHz H$_2$O line emission and the chance lensing configuration we propose suggest that spatially resolving the source emission may be difficult. On the other hand, we find that the CO~(11--10) emission probes similarly small physical scales and is significantly brighter than the 380~GHz H$_2$O emission. Observations of CO~(11--10) (or even higher excitation CO lines), combining the spatial resolution possible with the long baselines of ALMA and a high gravitational lens magnification, will present a promising approach to spatially resolve molecular tori at high redshift. This could allow detailed studies of supermassive black hole accretion and feedback on 10-pc-scales at the cosmic peak of black hole and galaxy growth.

\section*{Acknowledgements}

JPM acknowledges support from an NWO-CAS grant (project number 629.001.023).

This paper makes use of the following ALMA data: \\ ADS/JAO.ALMA\#2017.1.00316.S, ADS/JAO.ALMA\#2013.1.01110.S. \\ ALMA is a partnership of ESO (representing its member states), NSF (USA) and NINS (Japan), together with NRC (Canada), MOST and ASIAA (Taiwan), and KASI (Republic of Korea), in cooperation with the Republic of Chile. The Joint ALMA Observatory is operated by ESO, AUI/NRAO and NAOJ. 

The European VLBI Network is a joint facility of independent European, African, Asian, and North American radio astronomy institutes. Scientific results from data presented in this publication are derived from the following EVN project code(s): GW019. The National Radio Astronomy Observatory is a facility of the National Science Foundation operated under cooperative agreement by Associated Universities, Inc. 

This research made use of Astropy and Matplotlib packages for Python \citep{Astropy:2018,Hunter:2007}.

\bibliographystyle{mnras}
\bibliography{references}

\bsp	

\label{lastpage}
\end{document}